\documentclass[onecolumn,draftclsnofoot]{IEEEtran}
\def\isextended{1}

\ifdefined\isextended
	\newcommand{\appref}[1]{Appendix~\ref{#1}}
	\newcommand{\apprefall}{the appendices.}
\fi

\usepackage{amsmath,amsfonts,amssymb}
\usepackage{amsthm}
\usepackage{mathrsfs}
\usepackage{graphicx}
\usepackage{bm}

\input{niesen.def}

\graphicspath{ {./figures/} }

\begin{document}

\title{Energy-Efficiency Gains of Caching\\for Interference Channels}

\author{%
Jad~Hachem,
Urs~Niesen,
and~Suhas~Diggavi%
\thanks{This work was supported in part by NSF grants \#1423271 and \#1514531.}
}

\maketitle

\begin{abstract}
This paper initiates the study of energy-efficiency gains provided by caching.
We focus on the cache-aided Gaussian interference channel in the low-SNR regime.
We propose a strategy that creates content overlaps at the transmitter caches to allow for co-operation between the transmitters.
This co-operation yields a beamforming gain, which has to be traded off against a multicasting gain.
We evaluate the performance of this strategy and show its approximate optimality in both the single-receiver case and the single-transmitter case.

\end{abstract}

\section{Introduction}
\label{sec:intro}

The fundamental gains of caching were first derived for the error-free broadcast channel in \cite{maddah-ali2012}.
These consist of a local caching gain, which stems from the availability of a cache locally at each user, and a multicasting gain (also known as a global caching gain), which arises from the possibility of transmitting (coded) common information to multiple users.

The techniques developed in \cite{maddah-ali2012} take advantage of one aspect of the wireless medium: the broadcast of signals.
Another aspect of the wireless medium, which is not exploited in \cite{maddah-ali2012}, is the superposition of signals.
The wireless interference channel provides a setting that is suitable for the analysis of the gains of caching under both signal broadcast and superposition.
Recently, caching was studied for the Gaussian interference channel, with caches either at the transmitters \cite{maddah-ali2015interference,sengupta2016} or at both the transmitters and the receivers \cite{naderializadeh2016,xu2016,dof-caching}.
The focus in these works is on the high-SNR regimes, and the degrees-of-freedom gains of caching are by now well understood.

In this paper, we initiate the study of energy-efficiency gains of caching by considering a fast-fading Gaussian interference channel in the low-SNR regime with caches at transmitters and receivers.
We propose a separation-based strategy that uses the transmitter caches to enable a transmit beamforming gain in addition to the usual multicasting gain and local caching gain.
We find that there is a trade-off between the beamforming gain and the multicasting gain and propose two variants of the strategy, each of which prioritizes one of the two gains.
We show the approximate optimality (in the low-SNR regime) of each variant in two extreme cases: the variant prioritizing the beamforming gain is approximately optimal for the single-receiver case (i.e., the Gaussian multiple-access channel), while the variant prioritizing the multicasting gain is approximately optimal for the single-transmitter case (i.e., the Gaussian broadcast channel).

The remainder of this paper is organized as follows.
Section~\ref{sec:setup} formally describes the problem setting.
Section~\ref{sec:results} presents the main results of the paper.
The achievable strategy is described in detail in Section~\ref{sec:achievability}, and Section~\ref{sec:optim-mac} provides the proof of approximate optimality for the multiple-access case.
Proof details are relegated to \apprefall.

\section{Problem Setting}
\label{sec:setup}

A content library contains $N$ files, denoted by $W_1$ through $W_N$, of size $F$ bits each.
The content library is separated from its end users by a Gaussian interference network, whose receivers act as the users.
Let $L$ denote the number of transmitters in the network and $K$ denote the number of receivers (i.e., users).
Each transmitter is equipped with a cache of size $M_tF$ bits, and each receiver is equipped with a cache of size $M_rF$ bits.
The goal is to utilize the caches to help transmit files requested by the receivers across the interference network.
Two special cases that we will consider later in the paper are the single-transmitter (broadcast) case with $L=1$ and the single-receiver (multiple-access) case with $K=1$.

The system operates in two phases.
First, a \emph{placement phase} occurs during which each cache is filled with some function of the files.
This is done before the user demands are known.
Second, a \emph{delivery phase} occurs during which the user demands are revealed: each user $k$ requests a file $W_{d_k}$, where $d_k\in\{1,\ldots,N\}$.
Each transmitter $\ell$ responds by sending a codeword $\mathbf{x}_\ell=(x_\ell(1),\ldots,x_\ell(T))$ of length $T$ through the interference network.
The codeword $\mathbf{x}_\ell$ depends only on the user demands and the contents of transmitter $\ell$'s cache.
Receiver $k$ then observes at time $\tau$
\[
y_k(\tau) = \sum_{\ell=1}^L g_{k\ell}(\tau) x_\ell(\tau) + z_k(\tau),
\]
where $g_{k\ell}(\tau)$ are the i.i.d. complex channel gains, known causally at all transmitters and receivers, and $z_k(\tau)$ are i.i.d. additive white circulary-symmetric unit-variance complex Gaussian noise.
We assume the channel gains are uniform phase shifts, i.e., $g_{k\ell}(\tau) = e^{j\theta_{k\ell}(\tau)}$, where $j$ is the imaginary unit and $\theta_{k\ell}(\tau)$ are i.i.d. uniform over $[0,2\pi)$.
The channel inputs and outputs are also complex-valued.
Receiver $k$ then decodes its requested file from $\mathbf{y}_k$ and the contents of its cache.

We impose a power constraint of $P$ on the input, i.e.,
\[
||\mathbf{x}_\ell||^2 \le PT,\
\forall \ell\in\{1,\ldots,L\}.
\]
The \emph{rate} is defined as $R=F/T$.
For a given $P$, we wish to find the largest rate $R^*(P)$ such that, for all possible user requests $(d_1,\ldots,d_K)$,
\[
\max_k \Pr\left\{ \hat W_k \not= W_{d_k} \right\} \to 0\quad\text{as}\quad T\to\infty,
\]
where $\hat W_k$ denotes the reconstruction of file $W_{d_k}$ by user $k$.
In this paper we will focus on the capacity per unit energy~\cite{verdu}
\[
\widehat R^* = \lim_{P\to0^+} {R^*(P)}/{P}.
\]
This allows us to study the energy-efficiency gains that caching can provide.

\section{Main Results}
\label{sec:results}

Our main contribution is a separation-based communication strategy consisting of a physical layer and a network layer.
A message set is created from transmitters to receivers to serve as the interface between the physical layer and the network layer.
The physical layer transmits these messages across the interference network, while the network layer uses these messages as error-free bit pipes in order to deliver the requested files to the users.
This idea is similar to the one described in \cite{dof-caching} for the high-SNR regime.

It was shown in \cite{dof-caching} that, in the high-SNR regime, transmitter co-operation is not necessary for approximately achieving the degrees-of-freedom.
In contrast, in the low-SNR regime, transmitter co-operation becomes essential as it enables the transmit beamforming of signals to the receivers, yielding a power gain.
We therefore use the transmitter caches to create as much content overlap among the transmitters as possible, allowing them to co-operate and beamform signals to the intended receivers, thereby obtaining a significant power gain.
In general, we are able to obtain maximal multicasting (and local caching) gains, as well as a significant beamforming gain.
However, in special cases where the number of distinct file requests is small but the receiver memory is large, it is more beneficial to completely ignore the multicasting gain in favor of maximizing the beamforming gain.

In fact, there is a trade-off between the multicasting gain and the beamforming gain.
In order to obtain maximal multicasting gain, the receivers need to cache distinct parts of the files in order to increase the number of coding opportunities and thus enable the multicasting of coded messages.
Conversely, the beamforming gain can be improved by having all the receivers store common information.
This reduces the size of the total content that must be stored at the transmitters, which allows for greater overlap at the transmitters for the same memory size at the cost of losing the multicasting gain.

We therefore propose two different schemes, both of which utilize the separation-based approach: a multicasting scheme and a beamforming scheme.
The difference lies in the gain that each scheme prioritizes: the former prioritizes the multicasting (MC) gain while the latter prioritizes the beamforming (BF) gain.
Let $\widehat R_\text{MC}$ and $\widehat R_\text{BF}$ denote the bits per unit energy achieved by these schemes respectively.
By choosing the better of these two schemes in any given situation, we achieve
\begin{equation}
\label{eq:rate-max}
\widehat R^* \ge \max\left\{ \widehat R_\text{MC}, \widehat R_\text{BF} \right\}.
\end{equation}
The following two theorems provide the expressions for the bits per unit energy achieved by these schemes.

\begin{theorem}
\label{thm:rate-gc}
Let $\kappa=KM_r/N$ and $\lambda=LM_t/N$.
When $\kappa\in\{0,1,\ldots,K\}$ and $\lambda\in\{1,\ldots,L\}$, the multicasting scheme achieves
\[
\widehat R_\text{MC}
= \frac{1}{\ln2} \cdot
\frac{\kappa+1}{K-\kappa} \cdot \lambda
\cdot L.
\]
\end{theorem}

\begin{theorem}
\label{thm:rate-bf}
Let $\tilde\lambda=\min\{LM_t/(N-M_r),L\}$.
When $\tilde\lambda\in\{1,\ldots,L\}$, the beamforming scheme achieves
\[
\widehat R_\text{BF}
= \frac{1}{\ln2} \cdot
\frac{1}{\min\{N,K\}(1-M_r/N)} \cdot \tilde\lambda
\cdot L.
\]
\end{theorem}
Note that we abuse notation when $M_r=N$ (equivalently, $\kappa=K$), when we can achieve an arbitrarily large rate.

Theorems~\ref{thm:rate-gc} and~\ref{thm:rate-bf} give the rate achieved at specific corner points of the transmitter and receiver memories.
Since the \emph{inverse} of the rate is a convex function of $M_r$ and $M_t$ \cite{maddah-ali2015interference}, we can also achieve any linear combination of the inverse-rates of these points.

The next two subsections will analyze the two rate expressions and give a high-level overview of the schemes that achieve them.
At the end of the section, we discuss the approximate optimality of each scheme in special cases.

\subsection{The Multicasting Scheme}

The multicasting scheme prioritizes the multicasting gain.
To do so, it applies a receiver content placement strategy similar to the one in \cite{maddah-ali2012}, in which receivers store different content in a way that maximizes coding opportunities.
The transmitter content placement complements the receiver content placement by having subsets of transmitters share content.

More precisely, if $\kappa=KM_r/N$ and $\lambda=LM_t/N$ are integers, then every set of $\kappa$ receivers and $\lambda$ transmitters share some exclusive part of the content.
This creates opportunities for coded messages to be multicast to $\kappa+1$ receivers at a time \cite{maddah-ali2012} while simultaneously allowing every $\lambda$ transmitters to co-operate, beamform, and produce a power gain.

The result is then a maximized multicasting gain and a significant, though not necessarily maximized, beamforming gain.
More specifically, from Theorem~\ref{thm:rate-gc} the sum rate achieved by the multicasting scheme can be split into three components:
\begin{equation}
\label{eq:gains-gc}
K\widehat R_\text{MC}P
\approx \underbrace{\frac{1}{1-M_r/N}}_{G_\text{LC}}
\cdot \underbrace{\left(\frac{KM_r}{N}+1\right)}_{G_\text{MC}}
\cdot \underbrace{\frac{LM_t}{N}\vphantom{\left(\frac{KM_r}{N}\right)}}_{G_\text{BF}}
\cdot LP
\end{equation}
for $P$ small enough.
Here $G_\text{LC}$ is the local caching gain, $G_\text{MC}$ is the multicasting gain, and $G_\text{BF}$ is the beamforming gain.
In the equation, the $LP$ term can be thought of as the total power constraint on the transmitters.

Notice that the local caching gain and the multicasting (global caching) gain are at their maximal value.
Indeed, they are identical to those in \cite{maddah-ali2012}, whose setup consists of a single transmitter and an error-free broadcast link to all receivers.
The beamforming gain is approximately $LM_t/N$, which is equal to the number of copies of the content library that the transmitters can collectively store.
In the multicasting scheme, every subset of $LM_t/N$ transmitters share information in their caches, and they use this shared knowledge to co-operate and beamform messages to the receivers.
In a typical MISO channel, the beamforming gain is the number of co-operating antennas, and this is similar to $G_\text{BF}\approx LM_t/N$ in \eqref{eq:gains-gc}.

\subsection{The Beamforming Scheme}

The beamforming scheme ignores the multicasting gain in favor of improving the beamforming gain.
This is done by having all receivers store the exact same content in their caches and having transmitters co-operate and beamform the remaining part of the desired file individually to each receiver (no multicasting).
Since this makes a fraction of the content library available to all receivers, it is no longer necessary to store it at the transmitters.
This effectively reduces the size of the content library that is ``unavailable'' to the receivers---and hence that must be stored at the transmitters---down to $NF'=(N-M_r)F$ bits.
The transmitter memory can thus be expressed as $M_t/(1-M_r/N)\cdot F'$ bits.
Consequently, more overlap is made possible among the transmitters, thus increasing the beamforming gain to its maximal value.

This scheme is particularly useful when the number of receivers is smaller than the number of transmitters and the receiver memory is large compared to the transmitter memory.
In particular, it is approximately optimal when there is only one receiver, as discussed in Section~\ref{sec:results-approxoptim} below.

From Theorem~\ref{thm:rate-bf} we can write the sum rate of the beamforming scheme approximately as
\begin{equation}
\label{eq:gains-bf}
\widetilde K \widehat R_\text{BF}P
\approx \underbrace{\frac{1}{1-M_r/N}}_{G_\text{LC}}
\cdot \underbrace{\min\left\{\frac{LM_t/N}{1-M_r/N},L\right\}}_{G_\text{BF}}
\cdot LP
\end{equation}
for $P$ small enough, where $\widetilde K=\min\{N,K\}$ is the worst-case number of \emph{distinct} file requests.
Here $G_\text{LC}$ is the local caching gain and $G_\text{BF}$ is the beamforming gain.
Note the absence of any multicasting gain.
In the equation, the $LP$ term can again be thought of as the total power constraint on the transmitters.

Note that, when $M_t<N-M_r$, the expression $1-M_r/N$ normally associated with the local caching gain appears squared.
This is due to the double effect of a receiver's local cache: on the one hand it provides the local caching benefit to each receiver; on the other hand it reduces the size of the part of the library ``unavailable'' to the receivers by a factor of $1-M_r/N$, thus allowing for greater content overlaps among the transmitters.
Indeed, instead of sharing content between only $\lambda=LM_t/N$ transmitters, we can now increase this number to $\tilde\lambda=\min\{LM_t/(N-M_r),L\}\ge\lambda$, which explains the beamforming gain $G_\text{BF}$ in \eqref{eq:gains-bf}.

\subsection{Approximate Optimality}
\label{sec:results-approxoptim}

The following theorems state that our separation-based approach is approximately optimal in the low-SNR regime for two cases: the multiple-access case ($K=1$) and the broadcast case ($L=1$).
While the proof of approximate optimality for the broadcast case is a straightforward adaptation of the converse proof of \cite{maddah-ali2012} to the Gaussian low-SNR setup, the converse proof for the multiple-access case is more involved as it needs to capture the limits of possible co-operation among subsets of transmitters.

\begin{theorem}
\label{thm:optim-bc}
In the broadcast case, i.e., when $L=1$ and $M_t=N$, the bits per unit energy achieved by the multicasting scheme are approximately optimal,
\[
1 \le {\widehat R^*}/{\widehat R_\text{MC}} \le 12,
\]
for all $N\ge K$ and $M_r\in[0,N]$.%
\footnote{The case $N<K$ is handled in \appref{app:optim-bc}.}
\end{theorem}

The constant in Theorem~\ref{thm:optim-bc} can be numerically sharpened to about $8.151$ for $N,K\le100$.

\begin{theorem}
\label{thm:optim-mac}
In the multiple-access case, i.e., when $K=1$, the bits per unit energy achieved by the beamforming scheme are approximately optimal,
\[
1 \le {\widehat R^*}/{\widehat R_\text{BF}} \le 64,
\]
for all $N$, $L$, $M_r\in[0,N]$, and $M_t\in[(N-M_r)/L,N]$.
\end{theorem}

The constant in Theorem~\ref{thm:optim-mac} can be numerically sharpened to about $4.701$ for $N,L\le100$.
Note that Theorem~\ref{thm:optim-mac} holds for the entire memory regime of interest.

Notice that, in both these cases, we can assume without loss of generality that all the channel gains are one, i.e., all channel phase shifts are zero.
Indeed, when $K=1$, each transmitter can multiply its transmitted signal by the appropriate phase shift without affecting the power constraint or the (circularly symmetric) receiver noise.
Similarly, when $L=1$, each receiver can multiply its received signal by the appropriate phase shift.
For this reason, Theorems~\ref{thm:optim-bc} and~\ref{thm:optim-mac} apply for both fading and static channels.

Finally, we conjecture that our separation-based approach is approximately optimal in the low-SNR regime for fading channels for all values of $K$ and $L$, and proving this is part of our on-going work.

\subsection{Comparison with the High-SNR Regime}

We show in this paper that, in the low-SNR regime, caching can provide three gains: the local caching gain, the multicasting (global caching) gain, and the beamforming gain.
In the high-SNR regime, the first two gains are present, but instead of a beamforming gain there is an interference-alignment gain \cite{dof-caching}.
Notably, the interference-alignment gain does not require transmitter co-operation for approximate optimality, contrary to the beamforming gain in the low-SNR regime.
An interesting open problem is hence to analyze cache-aided communication in the transition regime from low to high SNR.

\section{Achievable Strategy}
\label{sec:achievability}

We adopt a separation-based strategy as discussed in Section~\ref{sec:results}, separating the network layer from the physical layer.
The idea is to create a set $\mathscr{V}$ of messages from (subsets of) transmitters and intended for (subsets of) receivers.
This message set acts as an interface between the network and physical layers: the physical layer transmits the messages across the interference channel, while the network layer uses them as error-free bit pipes in order to apply a caching strategy that delivers to each receiver its requested file.

Define $[m]=\{1,\ldots,m\}$.
Because of the symmetry in the problem, we will always choose message sets of the form
\begin{equation}
\label{eq:message-set}
\mathscr{V}_{pq}
\defeq \left\{ V_{\mathcal{K}\mathcal{L}} :
\mathcal{K} \subseteq [K], |\mathcal{K}|=p, \mathcal{L} \subseteq [L], |\mathcal{L}| = q
\right\},
\end{equation}
for some integers $p\in[K]$ and $q\in[L]$, where message $V_{\mathcal{K}\mathcal{L}}$ is to be sent collectively from the transmitters in $\mathcal{L}$ to the receivers in $\mathcal{K}$.
In other words, the messages are always from every subset of $q$ transmitters to every subset of $p$ receivers, for some $p$, $q$.
The physical layer assumes that message $V_{\mathcal{K}\mathcal{L}}$ is known to all the transmitters in $\mathcal{L}$.
At the network layer, we therefore need to ensure that any bits sent through the bit pipe represented by $V_{\mathcal{K}\mathcal{L}}$ are shared by all the transmitters in~$\mathcal{L}$.

Suppose that the physical layer is able to transmit all the messages in $\mathscr{V}_{pq}$ at a rate of $R'_{pq}$ each.
Suppose also that the network layer can send a total of $v_{pq}F$ bits through the messages (as bit pipes) in order to achieve its goal of delivering every file to the user that requested it.
Thus we have $R'_{pq}T=v_{pq}F$.
Since we also have $R=F/T$ by definition, this implies
\begin{equation}
\ifdefined\isextended
\else
\nonumber
\fi
\label{eq:net-phy-relation}
v_{pq}RT = R'_{pq}T \implies
R = R'_{pq}/v_{pq}.
\end{equation}
Therefore, by finding achievable values for $v_{pq}$ and $R'_{pq}$ for some pair $(p,q)$, we obtain an achievable rate $R$.

As previously mentioned, we propose two different schemes, the multicasting scheme and the beamforming scheme.
The difference in the two schemes lies in the network-layer strategy and the choice of~$p$ and~$q$: the multicasting scheme chooses to maximize $p$, whereas the beamforming scheme opts for maximizing $q$ and setting $p=1$.
The physical-layer strategy however is agnostic to the choice of schemes.

The physical-layer strategy is described below and in \appref{app:beamforming} along with its achieved rate $R'_{pq}$.
The network-layer strategies of the two schemes are provided in \appref{app:network} along with their achieved values of $v_{pq}$.

\subsection*{Physical-Layer Strategy}

Fix $p\in[K]$ and $q\in[L]$.
We wish to transmit the messages $\mathscr{V}_{pq}$ across the network.
Since we are focusing on the low-SNR regime, our strategy will attempt to get the largest power gain.

Consider a specific message $V_{\mathcal{K}\mathcal{L}}\in\mathscr{V}_{pq}$.
Since the transmitters in $\mathcal{L}$ all share the message $V_{\mathcal{K}\mathcal{L}}$, they can co-operate and beamform it to at least one user.
The idea is to schedule this message transmission when the channel is ``favorable'' for all the receivers in $\mathcal{K}$, at which point the transmitters can beamform to all receivers in $\mathcal{K}$ at once.
By ``favorable'', we mean that all the receivers in $\mathcal{K}$ can get approximately the maximum benefit (power gain) from this beamforming.
The result is the following achievable rate, proved in \appref{app:beamforming} where we describe the strategy in greater detail.

\begin{lemma}
\label{lemma:physical}
The message set $\mathscr{V}_{pq}$ can be transmitted across the interference network at a sum rate of
\[
\binom{L}{q}\binom{K}{p} \widehat{R}'_{pq}
\ge \frac{Lq}{\ln2}
\]
bits per unit energy, where $\widehat{R}'_{pq}=\lim_{P\to0^+}R'_{pq}(P)/P$.
\end{lemma}


\section{Approximate Optimality for the Multiple-Access Case}
\label{sec:optim-mac}

Recall that $K=1$ in this case.
Also recall that we can assume without loss of generality that all the channel gains are one.
In order to prove approximate optimality, we first derive the following cut-set bounds on the optimal rate.

\begin{lemma}
\label{lemma:cutset-mac}
For a single receiver (i.e., $K=1$), the optimal rate must satisfy
\[
R^*(P)
\le \max_{\substack{\mathbf{Q}\in\mathbb{C}^{L\times L}\\\mathbf{Q}\succeq0,\ Q_{\ell\ell}\le P}}
\,
\min_{\substack{\mathcal{L}\subseteq\{1,\ldots,L\}\\(L-|\mathcal{L}|)M_t<N-M_r}}
\frac{\log_2\left( 1 + \mathbf{1}^\top\mathbf{Q}_{\mathcal{L}|\mathcal{L}^c}\mathbf{1} \right)}
     {1 - \frac{M_r+(L-|\mathcal{L}|)M_t}{N}},
\]
where $\mathbf{1}$ is the all-ones vector, and
\[
\mathbf{Q}_{\mathcal{L}|\mathcal{L}^c}
=
\mathbf{Q}_{\mathcal{L},\mathcal{L}}
- \mathbf{Q}_{\mathcal{L},\mathcal{L}^c}
  \mathbf{Q}_{\mathcal{L}^c,\mathcal{L}^c}^{-1}
  \mathbf{Q}_{\mathcal{L}^c,\mathcal{L}}.
\]
\end{lemma}

We will now use Lemma~\ref{lemma:cutset-mac}, proved in \appref{app:optim-mac}, to prove Theorem~\ref{thm:optim-mac}, following a similar approach to \cite{niesen-diamond}.
The main idea is to use properties of the objective function of the maximization in Lemma~\ref{lemma:cutset-mac} to show that one maximizing covariance matrix $\mathbf{Q}$ has a symmetric structure, thereby reducing the maximization to just a single scalar variable.

We first swap the $\max$ over the covariance matrix $\mathbf{Q}$ and the $\min$ over the \emph{size} of the subset $\mathcal{L}$, giving
\[
R^*(P)
\le \min_{\substack{t\in[L]\\M_r+(L-t)M_t<N}}
\frac{N}{N-M_r-(L-t)M_t}
\max_{\mathbf{Q}}
\phi_t(\mathbf{Q}),
\]
where we have defined
\[
\phi_t(\mathbf{Q}) = \min_{|\mathcal{L}|=t}
\log_2\left(1+\mathbf{1}^\top\mathbf{Q}_{\mathcal{L}|\mathcal{L}^c}\mathbf{1} \right).
\]

By noticing that $\phi_t(\cdot)$ is both concave and invariant under permutation, we show in \appref{app:optim-mac} that one covariance matrix that maximizes $\phi_t(\cdot)$ must have the form
\begin{equation}
\ifdefined\isextended
\else
\nonumber
\fi
\label{eq:Q-form}
\mathbf{Q} = \left( (1-\rho)\mathbf{I} + \rho\mathbf{1}\mathbf{1}^\top \right) \cdot P
\end{equation}
for some $\rho\in[-1/(L-1),1]$.

We can now rewrite the upper bound on $R^*(P)$ as
\begin{equation}
\ifdefined\isextended
\else
\nonumber
\fi
\label{eq:rate-upper-bound}
\min_{\substack{t\in[L]\\L-t<\frac{N-M_r}{M_t}}}
\max_{\rho\in[\frac{-1}{L-1},1]}
\frac{t\left( 1 + (t-1)\rho - \frac{t(L-t)\rho^2}{1+(L-t-1)\rho} \right)}
{\left( 1 - \frac{M_r+(L-t)M_t}{N} \right)(\ln 2)} P,
\end{equation}
using $\log_2(1+x)\le x/\ln2$ and after some algebra.
By optimizing over $\rho$ and $t$, we obtain the result of the theorem.
For lack of space, we relegate this to \appref{app:optim-mac}.

\appendices
\section{Network-Layer Scheme
(Proof of Theorems~\ref{thm:rate-gc} and~\ref{thm:rate-bf})}
\label{app:network}

\begin{figure}
\centering
\includegraphics[width=.4\textwidth]{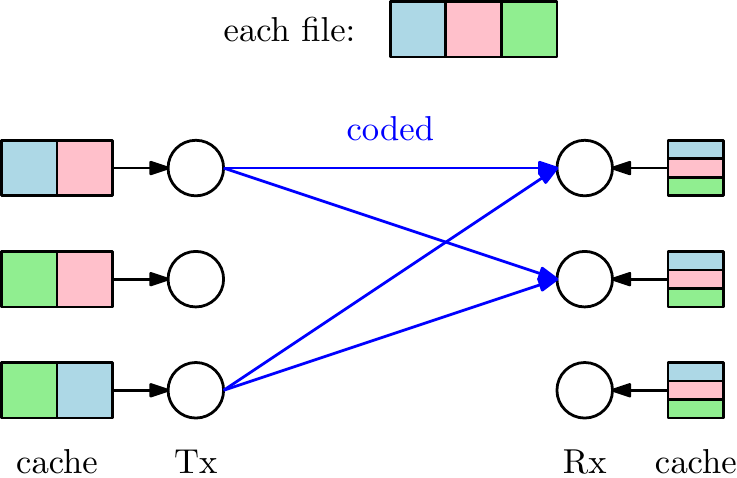}
\caption{An illustration of the multicasting scheme (only one file is shown for illustration), when $K=L=3$, $M_t=2N/3$, and $M_r=N/3$.
The multicasting scheme chooses $p=q=2$.
Each file is split into three subfiles, blue, pink, and green.
Every pair of transmitters caches one of the subfiles completely.
The receivers store each of the three subfiles according to the placement of \cite{maddah-ali2012}.
During the delivery phase, pairs of transmitters beamform a coded message to two receivers.}
\label{fig:multicasting}
\end{figure}

\begin{figure}
\centering
\includegraphics[width=.4\textwidth]{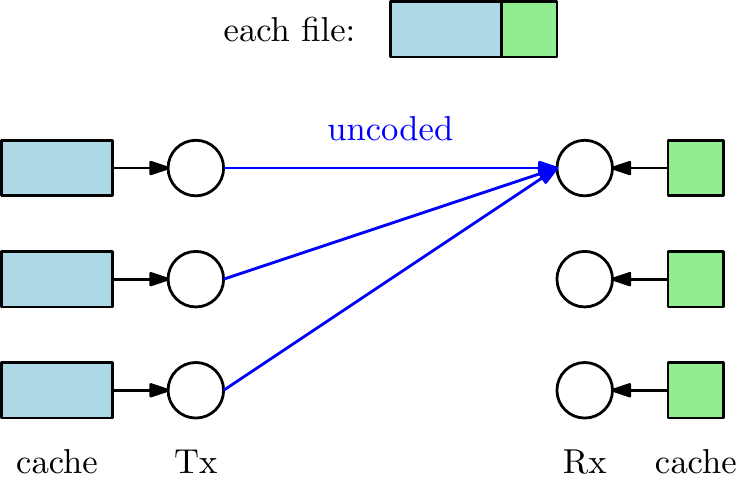}
\caption{An illustration of the beamforming scheme (only one file is shown for illustration), when $K=L=3$, $M_t=2N/3$, and $M_r=N/3$.
The beamforming scheme chooses $p=1$ and $q=3$.
Each file is split into two parts, blue and green.
Every receiver stores the green part completely.
In this example, all transmitters store the blue part completely (but in general they can store different parts).
During the delivery phase, all transmitters can beamform to send one uncoded message for each receiver.}
\label{fig:beamforming}
\end{figure}

In this appendix, we provide the details of the two network-layer strategies: the multicasting scheme and the beamforming scheme, illustrated in \figurename~\ref{fig:multicasting} and \figurename~\ref{fig:beamforming}, respectively.
This includes choosing $p$ and $q$ and determining the corresponding value of $v_{pq}$ that each scheme achieves, as introduced in Section~\ref{sec:achievability}.
Combined with Lemma~\ref{lemma:physical}, these imply the achievable rate results in Theorems~\ref{thm:rate-gc} and~\ref{thm:rate-bf}.

\subsection{Network-Layer Strategy: The Multicasting Scheme (Proof of Theorem~\ref{thm:rate-gc})}

Suppose $\kappa=KM_r/N$ and $\lambda=LM_t/N$ are both integers.
Collectively, the transmitters can hold $\lambda$ copies of the entire content library.
To take advantage of that, we first split every file $W_n$ into $\binom{L}{\lambda}$ equal subfiles $\{W_{n,\mathcal{L}}\}_{\mathcal{L}}$, where the index $\mathcal{L}$ is over all subsets of transmitters of size $\lambda$.
We can thus create $\binom{L}{\lambda}$ sublibraries: the sublibrary indexed by $\mathcal{L}$ contains the subfile $W_{n,\mathcal{L}}$ of every file $W_n$.
For the transmitter content placement, every transmitter $\ell$ stores all complete sublibraries indexed by $\mathcal{L}$ such that $\ell\in\mathcal{L}$.
The result is that every subset of transmitters of size $\lambda$ shares exactly one sublibrary.

For the receiver content placement, we first split each receiver cache into $\binom{L}{\lambda}$ equal parts and dedicate each part to one sublibrary.
We have thus divided our original problem into $\binom{L}{\lambda}$ subproblems.
In each subproblem, a subset $\mathcal{L}$ of transmitters shares a full sublibrary of $N$ subfiles of size $\tilde F=F/\binom{L}{\lambda}$ each.
Each of the $K$ receivers is equipped with a cache of size $M_rF/\binom{L}{\lambda}=M_r\tilde F$ bits, equivalently $M_r$ subfiles.
Since $\kappa=KM_r/N$, we can apply the strategy from \cite{maddah-ali2012} on this subproblem, which requires that the transmitters send a common message to every subset $\mathcal{K}$ of size $\kappa+1$ receivers.
We can enable that by choosing the message set $\mathscr{V}_{pq}$ with $p=\kappa+1$ and $q=\lambda$.

Each message $V_{\mathcal{K}\mathcal{L}}\in\mathscr{V}_{pq}$ has size $v_{pq}F$ bits, which can be rewritten in terms of the subfile size $\tilde F$ as $v_{pq}F=\binom{L}{\lambda}v_{pq}\tilde F$ bits.
From \cite{maddah-ali2012}, we know that the total number of bits that each subproblem needs to transmit across the bit pipes is $(K-\kappa)/(\kappa+1)\cdot \tilde F$, shared equally among all the bit pipes.
Therefore, the total number of bits sent through the $\binom{K}{\kappa+1}$ messages of each subproblem is
\[
\binom{K}{\kappa+1}\binom{L}{\lambda}v_{pq}\tilde F
= \binom{K}{\kappa+1}v_{pq}F
= \frac{K-\kappa}{\kappa+1}\tilde F.
\]
Consequently, we achieve
\begin{equation}
\label{eq:network-load-gc}
v_{pq} = \frac{K-\kappa}{\kappa+1} \cdot \frac{1}{\binom{L}{\lambda}\binom{K}{\kappa+1}}
\end{equation}
at the network layer.
By combining \eqref{eq:network-load-gc} with \eqref{eq:net-phy-relation} and Lemma~\ref{lemma:physical}, we obtain the result of Theorem~\ref{thm:rate-gc} for $\kappa$ and $\lambda$ integers.

\subsection{Network-Layer Strategy: The Beamforming Scheme (Proof of Theorem~\ref{thm:rate-bf})}

Recall that the beamforming scheme is different from the multicasting scheme in that it completely ignores any possible multicasting gain in favor of a larger beamforming gain.

Suppose $\tilde\lambda=\min\{LM_t/(N-M_r),L\}$ is an integer.
The first step is to divide each file $W_n$ into $\binom{L}{\tilde\lambda}+1$ parts,
\[
W_n = \left( W_{n,0}, W_{n,\mathcal{L}} : \mathcal{L}\subseteq[L], |\mathcal{L}|=\tilde\lambda \right),
\]
such that $W_{n,0}$ has size $M_rF/N$ bits and $W_{n,\mathcal{L}}$ has size $(N-M_r)F/\binom{L}{\tilde\lambda}$ for all $\mathcal{L}$.

In the placement phase, every receiver stores $W_{n,0}$ for every $n$.
Thus all receivers have exactly the same side information in their caches.
Each transmitter $\ell$ stores all parts $W_{n,\mathcal{L}}$ such that $\ell\in\mathcal{L}$.
Note that this placement satisfies the memory constaints $M_r$ and $M_t$ on the receivers and transmitters respectively.

During the delivery phase, every subset $\mathcal{L}$ of transmitters will beamform to each user $k$ the part of its requested file that these transmitters share.
Therefore, the message set that we choose is $\mathscr{V}_{pq}$ with $p=1$ and $q=\tilde\lambda$, and if user $k$ requests file $W_{d_k}$ then we set $V_{\{k\}\mathcal{L}}=W_{d_k,\mathcal{L}}$ for all $\mathcal{L}$.
Each message will as a result have a size of $v_{pq} = (N-M_r)/\binom{L}{\tilde\lambda}$.
Substituting in \eqref{eq:net-phy-relation} and using Lemma~\ref{lemma:physical}, we obtain the rate achieved in Theorem~\ref{thm:rate-bf}.

\section{Physical-Layer Scheme (Proof of Lemma~\ref{lemma:physical})}
\label{app:beamforming}

Recall that we wish to transmit the messages $\mathscr{V}_{pq}$ from \eqref{eq:message-set} across the interference network, for some $p\in[K]$ and $q\in[L]$.
As previously mentioned, the idea is to wait until a ``favorable'' channel occurs that allows some subset of transmitters to efficiently beamform some message to all its intended receivers at once.
In this proof, we focus on a particular $p$ and a particular $q$.

Let us focus on one subset pair $(\mathcal{K},\mathcal{L})$, where $\mathcal{K}$ is a subset of $p$ receivers and $\mathcal{L}$ is a subset of $q$ transmitters.
The most ``favorable'' channel to beamform message $V_{\mathcal{K}\mathcal{L}}$ occurs when the channel gains from the transmitters in $\mathcal{L}$ to each receiver in $\mathcal{K}$ are identical up to a multiplication by a scalar.
To be precise, the channel vectors $\mathbf{g}_{k\mathcal{L}}=(g_{k\ell})_{\ell\in\mathcal{L}}$ have to be equal for all $k\in\mathcal{K}$, up to a multiplication by a scalar.
However, since there are uncountably many values for each gain, the set of perfect channels has a measure of zero.
For this reason, we choose to divide the possible values of the channel gains into a finite number of bins $\beta\ge8$.

We will divide this proof into three parts: the first part presents the binning strategy, the second part gives the beamforming strategy and the corresponding analysis, and the third part analyzes the duty cycle, i.e., the fraction of time during which the channel is ``favorable'' for some transmitters and receivers.

\subsection{Binning strategy}

\begin{figure}
\centering
\includegraphics{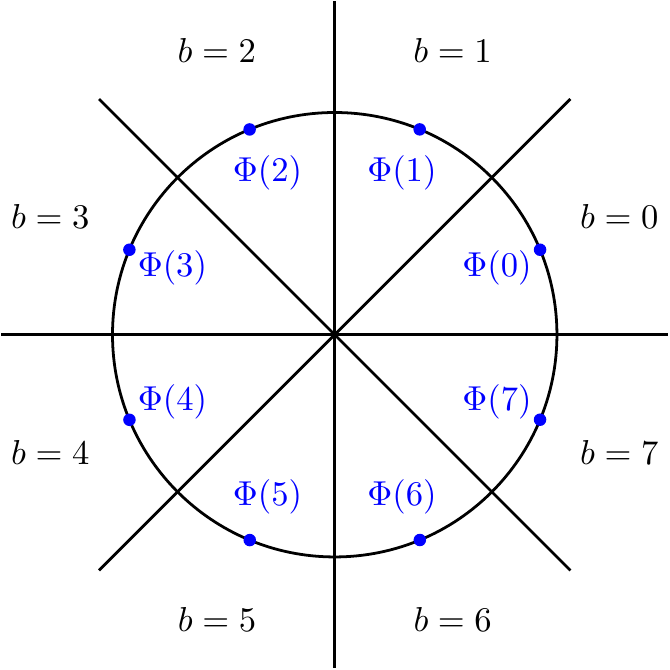}
\caption{The $\beta=8$ bins and their representative phases $\Phi(b)$.}
\label{fig:bins}
\end{figure}

Recall that the channel gains are phase shifts, $g_{k\ell}(\tau)=e^{j\theta_{k\ell}(\tau)}$, where $\theta_{k\ell}(\tau)\in[0,2\pi)$ uniformly.
For any angle $\theta\in[0,2\pi)$, define the binning function $B(\theta)$ as the unique integer such that
\[
\theta - \frac{2\pi}{\beta} B(\theta) \in [0,2\pi/\beta).
\]
Note that $B(\theta)\in\{0,\ldots,\beta-1\}$.
For each bin $b$, we define the representative phase of $b$ as the midpoint of all phases that are binned to $b$, i.e.,
\[
\Phi(b) = b \cdot 2\pi/\beta + \pi/\beta.
\]
This implies that $|\Phi(B(\theta))-\theta| \le \pi/\beta$ for all $\theta\in[0,2\pi)$.
The above-described binning is illustrated in \figurename~\ref{fig:bins} for a choice of $\beta=8$.
For simplicity, we will define $b_{k\ell}(\tau)=B(\theta_{k\ell}(\tau))$ to be the bin of the channel phase shift $\theta_{k\ell}(\tau)$ and $\phi_{k\ell}(\tau)=\Phi(b_{k\ell}(\tau))$ to be its representative phase.

We use these bins to determine which channels are ``favorable'' for a subset pair $(\mathcal{K},\mathcal{L})$.
Specifically, we say that a channel is favorable for $(\mathcal{K},\mathcal{L})$ if the corresponding channel vectors can be mapped to the same bins.
More formally, we say that the channel at time $\tau$ is \emph{favorable for $(\mathcal{K},\mathcal{L})$} if
\[
b_{k\ell}(\tau) = b_{k'\ell}(\tau)\quad \forall k,k'\in\mathcal{K},\forall \ell\in\mathcal{L}.
\]
We define $f_{\mathcal{K},\mathcal{L}}(\tau)$ to be one if the channel is favorable for $(\mathcal{K},\mathcal{L})$ at time $\tau$, and zero otherwise.
For every time $\tau$, we then define the set of pairs
\[
\mathscr{B}(\tau) = \left\{
(\mathcal{K},\mathcal{L}) : |\mathcal{K}|=p, |\mathcal{L}|=q, f_{\mathcal{K},\mathcal{L}}(\tau)=1
\right\}
\]
for which the channel is favorable.

\subsection{Beamforming strategy}

First, we encode each message $V_{\mathcal{K}\mathcal{L}}$ into a codeword $\mathbf{v}_{\mathcal{K}\mathcal{L}}$.
For every time $\tau$, we want to choose a pair $(\mathcal{K},\mathcal{L})$ for which the channel is favorable, if any exist.
We denote this pair by $(\mathcal{K}(\tau),\mathcal{L}(\tau))$, but we will ignore the $\tau$ index when it is obvious from context for clarity.
We then let the transmitters in $\mathcal{L}$ beamform a symbol $v_{\mathcal{K}\mathcal{L}}(\tau)$ from $\mathbf{v}_{\mathcal{K}\mathcal{L}}$ to the receivers in $\mathcal{K}$.

More formally, write $\mathcal{L}=\{\ell_1,\ldots,\ell_q\}$.
Let $\hat{\mathbf{b}}(\tau) = (\hat b_{\ell_1}(\tau),\ldots,\hat b_{\ell_q}(\tau))$ denote the vector of bins that resulted in the choice of subset pair at time $\tau$, i.e., $\hat b_{\ell}(\tau) = b_{k\ell}(\tau)$ for all $k\in\mathcal{K}$ and $\ell\in\mathcal{L}$.
Then, each transmitter $\ell\in\mathcal{L}$ sends
\[
x_\ell(\tau) = v_{\mathcal{K}\mathcal{L}}(\tau) \cdot e^{-j\Phi(\hat b_\ell(\tau))},
\]
and each receiver $k\in\mathcal{K}$ observes
\begin{IEEEeqnarray*}{rCl}
y_k(\tau)
&=& \sum_{\ell\in\mathcal{L}} e^{j\theta_{k\ell}(\tau)} \cdot e^{-j\Phi(\hat b_\ell(\tau))} v_{\mathcal{K}\mathcal{L}}(\tau) + z_k(\tau)\\
&=& v_{\mathcal{K}\mathcal{L}}(\tau) \sum_{\ell\in\mathcal{L}}
e^{j\left( \theta_{k\ell}(\tau) - \Phi(B(\theta_{k\ell}(\tau))) \right)}
+ z_k(\tau).
\end{IEEEeqnarray*}
The receiver SNR is then
\[
|v_{\mathcal{K}\mathcal{L}}(\tau)|^2 \cdot \left| \sum_{\ell\in\mathcal{L}}
e^{j\left( \theta_{k\ell}(\tau) - \Phi(B(\theta_{k\ell}(\tau))) \right)} \right|^2.
\]
Because of the binning, we can find a good lower bound on the magnitude of the sum term.
Let $\delta_{k\ell}(\tau) = \theta_{k\ell}(\tau) - \Phi(B(\theta_{k\ell}(\tau)))$.
Then,
\begin{IEEEeqnarray*}{rCl}
\left| \sum_{\ell\in\mathcal{L}} e^{j\delta_{k\ell}(\tau)} \right|^2
&=& \left( \sum_{\ell\in\mathcal{L}} e^{j\delta_{k\ell}(\tau)} \right)
\left( \sum_{\ell\in\mathcal{L}} e^{-j\delta_{k\ell}(\tau)} \right)\\
&=& \sum_{\ell\in\mathcal{L}} \left( 1
+ 2\sum_{\ell'>\ell} \Re\left\{ e^{j(\delta_{k\ell}(\tau)-\delta_{k\ell'}(\tau))} \right\}
\right)\\
&=& \sum_{\ell\in\mathcal{L}} \left( 1
+ 2\sum_{\ell'>\ell} \cos(\delta_{k\ell}(\tau)-\delta_{k\ell'}(\tau))
\right).
\end{IEEEeqnarray*}
Because $\delta_{k\ell}(\tau)\in[-\pi/\beta,\pi/\beta)$, then
\[
\delta_{k\ell}(\tau)-\delta_{k\ell'}(\tau) \in [-2\pi/\beta,2\pi/\beta],
\]
and hence, since $\beta\ge8$,
\[
\cos\left(\delta_{k\ell}(\tau)-\delta_{k\ell'}(\tau)\right) \ge \cos\frac{2\pi}{\beta}.
\]
We can write $\cos2\pi/\beta=(1-\gamma)$ for some $\gamma>0$.
Consequently,
\[
\left| \sum_{\ell\in\mathcal{L}} e^{j\delta_{k\ell}(\tau)} \right|^2
\ge \sum_{\ell\in\mathcal{L}}\left( 1 + (q-1) (1-\gamma) \right)
\ge (1-\gamma)q^2.
\]

Supposing that $|v_{\mathcal{K}\mathcal{L}}(\tau)|^2=P'$, and assuming that $V_{\mathcal{K}\mathcal{L}}$ is being transmitted during a fraction $\alpha$ of the total block length, we conclude that we can achieve a rate of
\begin{equation}
\label{eq:single-message-rate}
R'_{pq} \ge \alpha \log_2\left( 1 + (1-\gamma)q^2 \cdot P' \right)
\end{equation}
for message $V_{\mathcal{K}\mathcal{L}}$.

\subsection{Duty cycle analysis and achievable rate}

As mentioned previously, our strategy needs to wait for time instants $\tau$ such that $\mathscr{B}(\tau)$ is not empty.
We refer to the expected fraction of time during which it is not empty as the \emph{duty cycle} $\eta$, defined as $\eta=\Pr\{\mathscr{B}\not=\emptyset\}$.

When selecting pairs $(\mathcal{K},\mathcal{L})\in\mathscr{B}(\tau)$, it is possible to ensure that all pairs are selected equally likely.
For instance, if multiple pairs are possible for a specific $\tau$, we can pick one of them uniformly at random.
Thus the duty cycle will be shared equally among all pairs, and the expected fraction of time that any one message is being transmitted is $\alpha=\eta/\binom{L}{q}\binom{K}{p}$.
Since each transmitter is active for exactly $\binom{L-1}{q-1}\binom{K}{p}$ pairs out of the $\binom{L}{q}\binom{K}{p}$ total, then every transmitter will be active for a fraction
\[
\eta \cdot \frac{q}{L}
\]
of the time in expectation.
Consequently, it can scale its power by $L/\eta q$ during its duty cycle, which means
\[
P' = \frac{L}{\eta q} P.
\]
By appealing to the law of large numbers, it then follows from \eqref{eq:single-message-rate} that the set $\mathscr{V}_{pq}$ can be transmitted at a sum rate of
\[
\binom{L}{q}\binom{K}{p} R'_{pq}
\ge \eta \cdot \log_2\left( 1 + \frac{(1-\gamma)Lq}{\eta} P \right).
\]
When $P\le\sigma\cdot\eta/(1-\gamma)Lq$ for some $\sigma>0$, we get
\begin{equation}
\label{eq:sum-rate-power}
\binom{L}{q}\binom{K}{p} R'_{pq}
\ge (1-\gamma)Lq \cdot\frac{\log_2(1+\sigma)}{\sigma} \cdot P,
\end{equation}
by using $x\in[0,x_0]\implies\log_2(1+x)\ge x\cdot\log_2(1+x_0)/x_0$ for any $x_0>0$.

All that remains is to find a lower bound on the duty cycle $\eta$, in order to get a sufficient condition for the critical power necessary for \eqref{eq:sum-rate-power} to hold.
Consider the probability that a single subset pair $(\mathcal{K},\mathcal{L})$
gets a favorable channel at time $\tau$.
Recall that a channel is favorable for this pair if
\[
b_{k\ell}(\tau)=b_{k'\ell}(\tau)
\]
for all $k, k' \in \mathcal{K}$ and $\ell \in \mathcal{L}$.
Without loss of generality, we can assume that $b_{k1}(\tau)=0$ for all receivers $k$ since each receiver can always multiply its channel output with the correct phase shift.
Therefore, the above happens at time $\tau$ with probability
\[
\Pr\left\{f_{\mathcal{K},\mathcal{L}}(\tau)=1\right\} = \beta^{-(p-1)(q-1)}.
\]
Consequently,
\begin{IEEEeqnarray*}{rCl}
\eta
&=& \Pr\left\{ \mathscr{B}\not=\emptyset \right\}\\
&=& \Pr\left\{ \exists(\mathcal{K},\mathcal{L}) : f_{\mathcal{K},\mathcal{L}}(\tau) = 1 \right\}\\
&\overset{(a)}{\ge}& \Pr\left\{f_{\mathcal{K}_0,\mathcal{L}_0}(\tau)=1\right\}\\
&=& \beta^{-(p-1)(q-1)},
\end{IEEEeqnarray*}
for some arbitrary pair $(\mathcal{K}_0,\mathcal{L}_0)$.
Note that the inequality $(a)$ is quite loose; in practice the duty cycle should be higher because of the possibility to schedule all the $\binom{L}{q}\binom{K}{p}$ messages, and thus the critical power required for this analysis is higher.

Using this in \eqref{eq:sum-rate-power}, we get that
\[
\binom{L}{q}\binom{K}{p} R'_{pq}
\ge (1-\gamma)Lq\cdot\frac{\log_2(1+\sigma)}{\sigma} \cdot P
\]
bits per channel use, whenever $P\le\beta^{-(p-1)(q-1)}\sigma/(1-\gamma)Lq$.

Since $1-\gamma=\cos2\pi/\beta$, we can make $\gamma$ arbitrarily small by increasing the number of bins $\beta$.
Similarly, we know that $\log_2(1+\sigma)/\sigma$ approaches $1/\ln2$ as $\sigma$ approaches zero.
Therefore, for any $\epsilon>0$, we can choose particular values of $\beta$ and $\sigma$ so that, for a small enough $P$,
\[
\binom{L}{q}\binom{K}{p} R'_{pq}
\ge (1-\epsilon)\cdot \frac{LqP}{\ln2}
\]
bits per channel use.
This concludes the proof of Lemma~\ref{lemma:physical}.

\section{Approximate Optimality for the Broadcast Case
(Proof of Theorem~\ref{thm:optim-bc})}
\label{app:optim-bc}

The statement of Theorem~\ref{thm:optim-bc} as presented in Section~\ref{sec:results} holds for $N\ge K$ for ease of exposition and for lack of space.
In this appendix, we prove the following stronger result.
\begin{lemma}
\label{lemma:optim-bc}
In the broadcast case, i.e., when $L=1$ and $M_t=N$, we have
\[
1 \le \frac{\widehat{R}^*}{\max\{\widehat{R}_\text{MC},\widehat{R}_\text{BF}\}} \le 12,
\]
for all $N$, $K$, and $M_r\in[0,N]$.
\end{lemma}
Note that Theorem~\ref{thm:optim-bc} follows immediately from Lemma~\ref{lemma:optim-bc} since $\widehat{R}_\text{MC}\ge\widehat{R}_\text{BF}$ when $L=1$ and $N\ge K$.

We now prove Lemma~\ref{lemma:optim-bc}.
As previously mentioned, the channel gains are assumed to be one without loss of generality.
This implies that all the channel outputs are statistically equivalent.

From Theorem~\ref{thm:rate-gc}, we know that we can achieve
\[
\widehat R_\text{MC} \ge \frac{\kappa+1}{K-\kappa} \cdot \frac{1}{\ln2}
\]
bits per unit energy, when $\kappa=KM_r/N$ is an integer.
Moreover, for completeness we use the beamforming scheme in the case $N<K$.
We know from Theorem~\ref{thm:rate-bf} that we can also achieve
\[
\widehat R_\text{BF} \ge \frac{1}{\min\{N,K\}(1-M_r/N)} \cdot \frac{1}{\ln2} \cdot P.
\]
Thus by choosing the scheme that achieves the higher bits per unit energy, we can achieve
\begin{equation}
\label{eq:achieve-bc}
\max\{\widehat R_\text{MC},\widehat R_\text{BF}\}
\ge \frac{\max\{\kappa+1,K/N\}}{K-\kappa} \cdot \frac{P}{\ln2},
\end{equation}
when $\kappa=KM_r/N$ is an integer.

The upper bound is as follows.
Let $s\in\{1,\ldots,K\}$.
Denote by $U_k$ the contents of the cache of user $k$.
We observe the system after $\floor{N/s}$ instances, such that users $1$ through $s$ request a new file in each instance.
Thus the total number of requested files will be $\tilde N=s\floor{N/s}$, labeled $W_1$ through $W_{\tilde N}$.
During instance $i\in\{1,\ldots,\floor{N/s}\}$, denote $\mathbf{x}_1^i$ and $\mathbf{y}_k^i$ the channel input of the transmitter and channel output of receiver $k$, respectively.

Consider now the caches $U_1,\ldots,U_s$ and the channel output $\mathbf{y}_1$.
Since all channel outputs are statistically equivalent, these are enough to decode anything that users $1$ through $s$ can decode.
Therefore,
\begin{IEEEeqnarray*}{rCl}
s\floor{N/s} RT
&=& s\floor{N/s} F\\
&=& H\left( W_1,\ldots,W_{\tilde N} \right)\\
&\overset{(a)}{\le}& I\left( W_1,\ldots,W_{\tilde N} ; U_1,\ldots,U_s, \mathbf{y}_1^1,\ldots,\mathbf{y}_1^{\floor{N/s}} \right)\\
&& {} + \epsilon T\\
&\le& I\left( W_1,\ldots,W_{\tilde N} ; \mathbf{y}_1^1,\ldots,\mathbf{y}_1^{\floor{N/s}} \right)\\
&& {} + H\left( U_1,\ldots,U_s \right)
+ \epsilon T\\
&\overset{(b)}{\le}& I\left( \mathbf{x}_1^1,\ldots,\mathbf{x}_1^{\floor{N/s}} ; \mathbf{y}_1^1,\ldots,\mathbf{y}_1^{\floor{N/s}} \right)\\
&& {} + H\left( U_1,\ldots,U_s \right)
+ \epsilon T\\
&\overset{(c)}{\le}& \floor{N/s} \cdot I\left( \mathbf{x}_1 ; \mathbf{y}_1 \right)
+ sM_rRT + \epsilon T\\
&\overset{(d)}{\le}& \floor{N/s} \cdot T\log_2\left( 1 + P \right)
+ sM_rRT + \epsilon T\\
&\overset{(e)}{\le}& \floor{N/s} \frac{P}{\ln2}T + sM_rRT + \epsilon T,
\end{IEEEeqnarray*}
where $(a)$ uses Fano's inequality, $(b)$ uses the data processing inequality, $(c)$ applies the memory constaints on the receiver caches, $(d)$ uses the capacity bound for a point-to-point Gaussian channel, and $(e)$ uses $\ln(1+x)\le x$.
Consequently,
\begin{equation}
\label{eq:converse-bc}
R^*(P) \le \min_{s\in\{1,\ldots,K\}} \frac{1}{s\left( 1 - M_r/\floor{N/s} \right)} \cdot \frac{P}{\ln2}.
\end{equation}

The upper and lower bounds in \eqref{eq:achieve-bc} and \eqref{eq:converse-bc} are identical to their analogues in \cite{maddah-ali2012}, up to a multiplicative constant.
Therefore, the same argument used in \cite{maddah-ali2012} proves that
\[
\frac{\widehat{R}^*}{\max\{\widehat{R}_\text{MC},\widehat{R}_\text{BF}\}} \le 12.
\]
This proves Lemma~\ref{lemma:optim-bc} and, by extension, Theorem~\ref{thm:optim-bc}.

\section{Approximate Optimality for the Single-Receiver Case
(Proof of Theorem~\ref{thm:optim-mac})}
\label{app:optim-mac}

First, we prove that there exists an \emph{optimal} covariance matrix $\tilde{\mathbf{Q}}$ of the form in \eqref{eq:Q-form}, using the two properties of $\phi_t$: concavity and invariance under permutation.

Let $\mathbf{Q}^*$ be a covariance matrix that maximizes $\phi_t$.
Define $\tilde{\mathbf{Q}} = \frac{1}{L!} \sum_{\bm{\pi}} \bm{\pi}^\top\mathbf{Q}^*\bm{\pi}$.
By the two properties of $\phi_t$, we have
\[
\phi_t(\tilde{\mathbf{Q}})
\overset{(a)}{\ge} \frac{1}{L!} \sum_{\bm{\pi}} \phi_t\left( \bm{\pi}^\top\mathbf{Q}^*\bm{\pi} \right)
\overset{(b)}{=} \phi_t(\mathbf{Q}^*),
\]
where $(a)$ uses concavity of $\phi_t$ and $(b)$ uses its invariance under permutation.
Therefore, $\tilde{\mathbf{Q}}$ also maximizes $\phi_t$.
Moreover, we can see that $\bm{\pi}^\top\tilde{\mathbf{Q}}\bm{\pi}=\tilde{\mathbf{Q}}$ for any permutation $\bm{\pi}$, which implies that $\tilde{\mathbf{Q}}$ must have the form
\[
\tilde{\mathbf{Q}} = \left( (1-\rho)\mathbf{I} + \rho\mathbf{1}\mathbf{1}^\top \right) \cdot P
\]
for some $\rho$.
In order for $\tilde{\mathbf{Q}}$ to be positive semidefinite, we need $\rho\in[-1/(L-1),1]$.

Using the structure of $\tilde{\mathbf{Q}}$, we can simplify the analysis to the following.
Recall from Section~\ref{sec:optim-mac} and \eqref{eq:rate-upper-bound} that this simplifies the upper bound on the optimal expected rate to
\begin{equation}
\label{eq:converse-t}
R^*(P) \le
\min_{\substack{t\in[L]\\(L-t)M_t+M_r<N}}
\frac{\Psi(t)}{1 - \frac{M_r+(L-t)M_t}{N}} \cdot\frac{P}{\ln2}
\end{equation}
bits per channel use, where
\[
\Psi(t) =
\max_{\rho\in[\frac{-1}{L-1},1]}
t\left( 1 + (t-1)\rho - \frac{t(L-t)\rho^2}{1+(L-t-1)\rho} \right).
\]
Let us start with the maximization over $\rho$.
We can focus on the function
\[
f(\rho) = (t-1)\rho - \frac{t(L-t)\rho^2}{1+(L-t-1)\rho},
\]
which is the only part that depends on $\rho$.
Differentiating $f$,
\begin{IEEEeqnarray*}{rCl}
f'(\rho)
&=& t-1\\
&& {} - \frac{2t(L-t)\rho\left(1+(L-t-1)\rho\right) - (L-t-1)t(L-t)\rho^2}
			{\left[1+(L-t-1)\rho\right]^2}\\
&=& t-1
- \frac{t(L-t)\rho\left( 2 + (L-t-1)\rho \right)}
	{\left[1+(L-t-1)\rho\right]^2}.
\end{IEEEeqnarray*}
The sign of $f'(\rho)$ is the same as the sign of
\begin{IEEEeqnarray*}{rCl}
g(\rho)
&=& (t-1)\left[ 1 + (L-t-1)\rho \right]^2
- t(L-t)\rho\left( 2 + (L-t-1)\rho \right)\\
&=& (t-1)\left( 1 + 2(L-t-1)\rho + (L-t-1)^2\rho^2 \right)\\
&& {} - t(L-t)\rho\left( 2 + (L-t-1)\rho \right)\\
&=& t-1 + 2(t-1)(L-t-1)\rho + (t-1)(L-t-1)^2\rho^2\\
&& {} - 2t(L-t)\rho - t(L-t)(L-t-1)\rho^2\\
&=& t - 1\\
&& {} + 2\left[ t(L-t)-t-(L-t)+1 - t(L-t) \right] \rho\\
&& {} +  \left[ (t-1)(L-t)^2 - 2(t-1)(L-t) + (t-1) \right.\\
&& \qquad \left. {} - t(L-t)^2 + t(L-t) \right] \rho^2\\
&=& t - 1 - 2(L-1)\rho\\
&& {} + \left[ -(L-t)^2 - (t-2)(L-t) + (t-1) \right] \rho^2\\
&=& t-1 - 2(L-1)\rho - (L-1)(L-t-1)\rho^2.
\end{IEEEeqnarray*}
Thus to find the maximum of~$f$ we first find the roots of~$g$.
If $t\not=L-1$, then $g(\rho)$ is a quadratic with discriminant $\Delta=4t(L-1)(L-t)$, which yields the roots
\[
\rho
= \frac{2(L-1) \pm 2\sqrt{t(L-1)(L-t)}}{-2(L-1)(L-t-1)}\\
= \frac{-1 \mp \sqrt{\frac{t(L-t)}{L-1}}}{L-t-1}.
\]
Therefore, in the range $\rho\in[-1/(L-1),1]$, the function $f(\rho)$ reaches a maximum when
\[
\rho^* = \frac{-1+\sqrt{t(L-t)/(L-1)}}{L-t-1}.
\]
The maximum is thus
\[
\max_{\rho\in[-1/(L-1),1]} f(\rho)
= f(\rho^*)
= \left[ \frac{\sqrt{t(L-t)}-\sqrt{L-1}}{L-t-1} \right]^2.
\]
If $t=L-1$, then $g(\rho)=0$ for $\rho=(L-2)/2(L-1)$, yielding
\[
f(\rho^*) = \frac{(L-2)^2}{4(L-1)}.
\]

We therefore get
\[
\Psi(t) = \begin{cases}
t\left( 1 + \left[ \frac{\sqrt{t(L-t)}-\sqrt{L-1}}{L-t-1} \right]^2 \right)
& \text{if $t\not=L-1$;}\\
L^2/4 & \text{if $t=L-1$.}
\end{cases}
\]

We will now complete the proof of Theorem~\ref{thm:optim-mac}.
Recall from Theorem~\ref{thm:rate-bf} that, for $K=1$ and for a small enough $P$, we can achieve
\[
\widehat R_\text{BF}
\ge \frac{1}{\ln2} \cdot \frac{L\tilde\lambda}{1-M_r/N} \cdot P
\]
bits per unit energy, when $\tilde\lambda = \min\{ LM_t/(N-M_r) , L \}$ is an integer.
For a general $\tilde\lambda$, we can lower-bound the rate at $\tilde\lambda$ by the rate at $\floor{\tilde\lambda}$, which yields
\begin{IEEEeqnarray*}{rCl}
\widehat R_\text{BF}
&\ge& \frac{1}{\ln2} \cdot \frac{L\floor{\tilde\lambda}}{1-M_r/N} \cdot P\\
&\overset{(a)}{\ge}& \frac{1}{2\ln2} \cdot \frac{L\tilde\lambda}{1-M_r/N} \cdot P,
\IEEEyesnumber\label{eq:gap-proof-ach}
\end{IEEEeqnarray*}
where $(a)$ is due to $\tilde\lambda\ge1$.

The rest of the proof is split into two cases: $M_t\ge(N-M_r)/4$ and $M_t<(N-M_r)/4$.

\subsubsection*{Case 1}
If $M_t\ge(N-M_r)/4$, then $\tilde\lambda\ge L/4$, and hence \eqref{eq:gap-proof-ach} gives
\begin{equation}
\label{eq:gpf-ach-1}
\widehat R_\text{BF} \ge \frac{1}{8\ln2} \cdot \frac{L^2}{1-M_r/N} \cdot P.
\end{equation}

Choosing $t=L$, which satisfies the condition $(L-t)M_t+M_r<N$, in \eqref{eq:converse-t}, we get $\Psi(L)=L^2$, yielding the upper bound on the optimal rate
\begin{equation}
\label{eq:gpf-conv-1}
R^*(P)
\le \frac{L^2}{1-M_r/N} \cdot \frac{P}{\ln2}.
\end{equation}
Combining \eqref{eq:gpf-ach-1} with \eqref{eq:gpf-conv-1}, we get
\begin{equation}
\label{eq:gpf-gap-1}
\frac{\widehat R^*}{\widehat R_\text{BF}}
\le 8.
\end{equation}

\subsubsection*{Case 2}
If $M_t<(N-M_r)/4$, then $\tilde\lambda = LM_t/(N-M_r)$ and \eqref{eq:gap-proof-ach} becomes
\begin{equation}
\label{eq:gpf-ach-2}
\widehat R_\text{BF} \ge \frac{1}{2\ln2} \cdot \frac{L^2M_t/N}{(1-M_r/N)^2} \cdot P.
\end{equation}
We apply \eqref{eq:converse-t} using
\[
t = L - \left\lfloor \frac{N-M_r}{2M_t} \right\rfloor.
\]
This satisfies the condition $(L-t)M_t+M_r<N$.
Furthermore, it implies $t\le L-2$.

The denominator of \eqref{eq:converse-t} can be lower-bounded by
\[
1 - \frac{M_r+(L-t)M_t}{N} \ge \frac12 \left( 1 - \frac{M_r}{N} \right),
\]
which implies
\[
R^*(P)
\le \frac{\Psi(t)}{\frac12(1-M_r/N)} \cdot \frac{P}{\ln2}.
\]
Because $t\ge1$ and $t\le L-2$, we can upper-bound $\Psi(t)$ by
\begin{IEEEeqnarray*}{rCl}
\Psi(t)
&=& t \left( 1 + \left[ \frac{\sqrt{t(L-t)}-\sqrt{L-1}}{L-t-1} \right]^2 \right)\\
&\overset{(a)}{\le}& L \left( 1 + \frac{t(L-t)}{(L-t)^2(1-\frac{1}{L-t})^2} \right)\\
&\le& L \left( 1 + \frac{4t}{L-t} \right)\\
&=& L \left( 1 + 4\frac{L-\floor{(N-M_r)/2M_t}}{\floor{(N-M_r)/2M_t}} \right)\\
&=& L \left( 1 + \frac{4L}{\floor{(N-M_r)/2M_t}} - 4 \right)\\
&\le& \frac{4L^2}{\floor{(N-M_r)/2M_t}}\\
&\le& \frac{16L^2M_t}{N-M_r},
\end{IEEEeqnarray*}
where $(a)$ follows from the fact that $t(L-t)\ge L-1$ for all $t\in[1,L-1]$.
Therefore,
\begin{equation}
\label{eq:gpf-conv-2}
R^*(P) \le \frac{32L^2M_t/N}{(1-M_r/N)^2} \cdot \frac{P}{\ln2}.
\end{equation}

Combining \eqref{eq:gpf-ach-2} with \eqref{eq:gpf-conv-2}, we get
\begin{equation}
\label{eq:gpf-gap-2}
\frac{\widehat R^*}{\widehat R_\text{BF}}
\le 64.
\end{equation}

Together, \eqref{eq:gpf-gap-1} and \eqref{eq:gpf-gap-2} give the result of Theorem~\ref{thm:optim-mac}.

\begin{IEEEproof}[Proof of Lemma~\ref{lemma:cutset-mac}]
Recall that all channel gains are one without loss of generality.
We consider $N$ realizations of the problem, during each of which the user requests a new file.
When it requests file $W_n$, we denote the channel inputs by $\mathbf{x}_\ell^n$ and the channel output by $\mathbf{y}_1^n$.
Furthermore, let $U_1$ denote the cache of receiver $1$, and $V_\ell$ denote the cache of transmitter $\ell$.
\begin{IEEEeqnarray*}{rCl}
NRT
&=& NF\\
&=& H\left(W_1,\ldots,W_N\right)\\
&=& I\left(W_1,\ldots,W_N ; U_1,\mathbf{y}_1^1,\ldots,\mathbf{y}_1^N \right)\\
&& {} + H\left(W_1,\ldots,W_N \middle| U_1,\mathbf{y}_1^1,\ldots,\mathbf{y}_1^N \right)\\
&\overset{(a)}{\le}& I\left(W_1,\ldots,W_N ; U_1,\mathbf{y}_1^1,\ldots,\mathbf{y}_1^N \right)
+ \epsilon T\\
&\le& I\left(W_1,\ldots,W_N ; \mathbf{y}_1^1,\ldots,\mathbf{y}_1^N \right)
+ H(U_1) + \epsilon T\\
&\le& I\left(W_1,\ldots,W_N ; \mathbf{y}_1^1,\ldots,\mathbf{y}_1^N \middle| \mathbf{x}_{\mathcal{L}^c}^1,\ldots,\mathbf{x}_{\mathcal{L}^c}^N \right)\\
&& {} + I\left(W_1,\ldots,W_N ; \mathbf{x}_{\mathcal{L}^c}^1,\ldots,\mathbf{x}_{\mathcal{L}^c}^N \right)\\
&& {} + H\left( U_1 \right) + \epsilon T\\
&\overset{(b)}{\le}& I\left(\mathbf{x}_{\mathcal{L}}^1,\ldots,\mathbf{x}_{\mathcal{L}}^N ; \mathbf{y}_1^1,\ldots,\mathbf{y}_1^N \middle| \mathbf{x}_{\mathcal{L}^c}^1,\ldots,\mathbf{x}_{\mathcal{L}^c}^N \right)\\
&& {} + H\left( V_{\mathcal{L}^c} \right) + H\left( U_1 \right) + \epsilon T\\
&\overset{(c)}{\le}& N I\left(\mathbf{x}_{\mathcal{L}} ; \mathbf{y}_1 \middle| \mathbf{x}_{\mathcal{L}^c} \right)
+ (L-|\mathcal{L}|)M_tRT + M_rRT + \epsilon T\\
&\overset{(d)}{\le}& NT\log_2\left(1 + \mathbf{1}^\top\mathbf{Q}_{\mathcal{L}|\mathcal{L}^c}\mathbf{1} \right)\\
&& {} + (L-|\mathcal{L}|)M_tRT + M_rRT + \epsilon T,
\end{IEEEeqnarray*}
where $(a)$ uses Fano's inequality, $(b)$ follows from the data processing inequality, $(c)$ applies the memory constraints on the caches, and $(d)$ is the MISO channel bound.
\end{IEEEproof}

\bibliographystyle{IEEEtran}
\bibliography{journal,xchannel,caching}

\end{document}